# Towards interpreting ML-based automated malware detection models: a survey

Yuzhou Lin[1], Xiaolin Chang[1]

[1] Beijing Key Laboratory of Security and Privacy in Intelligent Transportation, Beijing Jiaotong University, China

**Abstract.** Malware is being increasingly threatening and malware detectors based on traditional signature-based analysis are no longer suitable for current malware detection. Recently, models based on machine learning (ML) are developed for predicting unknown malware variants and saving human strength. However, most of the existing ML models are black-box, which made their prediction results undependable, and therefore need further interpretation in order to be effectively deployed in the wild. This paper aims to examine and categorize the existing researches on ML-based malware detector interpretability. We first give a detailed comparison over the previous work on common ML model interpretability methods in groups after introducing the principles, attributes, evaluation indicators and taxonomy of common ML interpretability. Then we investigate the interpretation methods towards malware detection, by addressing the importance of interpreting malware detectors, challenges faced by this field, solutions for mitigating these challenges, and a new taxonomy for classifying all the state-of-the-art malware detection interpretability work in recent years. The highlight of our survey is providing a new taxonomy towards malware detection interpretation method based on the taxonomy summarized by previous researches in the common field. In addition, we are the first to evaluate the state-of-the-art approaches by interpretation method attributes to generate the final score so as to give insight to quantifying the interpretability. By concluding the results of the recent researches, we hope our work can provide suggestions for researchers who are interested in the interpretability on ML-based malware detection models.

## 1 Introduction

Malware detectors based on machine learning (ML) have been becoming more popular in malware detection because malware and its variants are being increasingly threatening. More than 800 million malwares have been reported by 2020 so far, and more than 65 million new malware variants were detected in 2019 as an increase of 35.4% compared with that of last year [3]. More important, there are more malware variants cannot be recognized by the known signatures than before. In general, according to McAfee latest report of 2019, ransomware attacks grew by 118%, PowerShell attacks increased by 460%, and more than 4 million unique sources of SMB exploit traffic have been found. The traditional signature-based approaches cannot be given up due to their high interpretability and dependable analysis results. Besides, ML-based methods can be used to predict unknown samples in wild but often lack interpretability. Here we give a quick look from signature-based methods to ML-based methods.

Although signature-based malware detectors can have high accuracy in discriminating malware, these methods are only able to recognize the known malicious applications. In order to predict the unknown samples whose features have not been stored in the signature space, ML-based detection is getting increasingly popular. Nevertheless, these approaches would require highly preprocessed features to train and learn for classifiers, which cost a lot of time and lack the intelligent analysis.

To solve the aforementioned limitations, Deep Learning (DL) was developed for malware detection, such as using Recurrent Neural Network (RNN) to improve the capability of latent semantic extraction in API-sequence-based malware classification, and using Convolutional Neural Network (CNN) to process CV problems generated from malware. DL approaches mainly involve feature vector representation, IMG-based representation, API call traces, instruction traces, bytes-based representation and network traffic, etc. Although these DL approaches always have high prediction accuracy, the interpretability is low and need some specific post hoc processing methods to get outcomes interpretable to some degree.

Although the detection efficiency and prediction accuracy have been improved in the past years, the interpretation of ML technology, especially DL, is still being an open challenge, and most of the models used today for training are treated as a black box model. The high prediction accuracy would be undependable if the model cannot be interpreted well. As far as we know, a black box will generate an output through a series of systematic processes which is hard for a human to understand. Therefore, many researchers turn back to traditional signature-based malware detection approaches but it is beyond doubt that these approaches cannot be suitable for today's malware analysis.

The major research challenges faced by ML-based malware detection are structured as problem groups like class imbalance, open and public benchmarks for research, the problem of concept drift and various adversarial learning techniques against ML-based detectors, interpretability of the models, etc. [1]. Among all, model interpretability has not been surveyed systematically. Therefore, our survey is going to investigate this challenge. Specifically, the main contributions of this paper are summarized as follows.

1) We use several interpretability evaluation indicators to compare and discuss the existing work in the common ML interpretability field being grouped by a given taxonomy from References.

2) We summarize the malware detection interpretability solution logic and provide a new taxonomy of interpretation methods towards malware detection. Based on this taxonomy, we then recapitulate the state-of-the-art interpretation work towards malware detection.

3) We define a novel evaluation calculation approach by giving different weights to interpretation method attributes to compare those state-of-the-art interpretation work on malware detection. This approach reduces the difficulty of quantifying the model interpretability.

4) We systematically summarize the present challenges faced by interpreting malware detectors and provide a potential future workflow in this field.

The rest of the article is organized as follows. Section 2 describes the concept of interpretability, attributes of interpretation methods, evaluation indicators towards common interpretation methods, and the classification of the existing interpretability work in the common field. Section 3 provides a detailed review and evaluation of the newest investigation on interpretation methods in the malware detection field. Section 4 presents a summary towards model interpretability based on malware detection and gives discussions about future work and challenges.

## 2 The interpretability of common ML-based models

In this section, we introduce the basic concept of interpretability, attributes, evaluation indicators and the classification of the existing common interpretation methods. Then, we compare and evaluate the interpretability of all these methods in groups by several indicators in order to help researchers choose a better solution in a specific case.

### 2.1 Description of interpretability

Each ML model has a response function, which is used to establish the relationship between independent variables and dependent variables. When a model makes a prediction, a common interpretation method tries to understand the decisions made by these response functions. Any interpretation of a ML model must consider at least three aspects: what, why and how.

- **What**. We can analyze the model to find out the potential feature interactions and understand which features are important in the decision-making strategy of the model.

- **Why**. We definitely need to verify why there exist some critical features that can be crucial for prediction and which features are key features to keep models dependable.

- **How**. We should know how to evaluate data points that affect the model prediction performance and how to use these data points to explore the model transparency.

### 2.2 Attributes of an interpretation method

Each interpretation method at least has the following four attributes, which can be used to evaluate the existing interpretation methods. We use these attributes in Section 3.4.13.

**Attribute1: Intrinsic or post hoc?** This attribute determines whether an interpretation method studies the inner structure of a model. Intrinsic interpretation methods refer to that we use some interpretable ML models (such as linear model, parameter model, tree model) to predict the unknown test samples and the outcomes can be interpreted by the models themselves. These methods can be used for interpreting models by analyzing the inner structure of the models and can always give global interpretation. Post hoc interpretability refers to the process after the training stage, and this kind of methods always focus on the relevance between the input features and the prediction results.

**Attribute2: Local or global?** Global interpretation means to explain the predicted behaviors of the whole model, while local interpretation is to explain a single input and output via a particular model. Their comparison is depicted in **Table 1** in terms of purpose, requirements and challenges.

**Table 1. Scope measures**

| | Purpose | Requirements | Challenges |
|---|---|---|---|
| Global interpretability | Interpret model decisions based on conditional interactions between dependent and independent variables on a complete data set | Complete data sets, complete model structure, assumptions and constraint knowledge | It is difficult to visualize features with more than two dimensions |
| Local interpretability | Understand why a single model makes such a decision based on a single instance | The data point and the existence of the local subarea in the feature space around the point | Local data distribution and feature space may behave completely different |

**Attribute3: Result of the interpretation methods**. Interpretation methods can have attributes according to their result type. These attributes include feature summary statistic, feature visualization, model internals like learned weights, data points like using counterfactual explanations, and intrinsically interpretable models like approximating black-box models with a surrogate model. In general, these methods give interpretation by looking at model internals or feature summaries. The detailed description and examples for these types are illustrated in **Table 2**. 'Principle' refers to the meaning of the attribute, and 'Example' refers to the specific methods with the corresponding attributes.

**Table 2 Interpretation method attributes according to the result of these methods**

| | Principle | Example. |
|---|---|---|
| Feature summary statistic | A process of summarizing the feature statistic in different formats. | A single number per feature to represent the feature importance and pairwise feature interaction strengths for a more complex result etc. |
| Feature summary visualization | A process of visualizing the feature summary by some plots. | The partial dependence plots (PDP), accumulated local effect (ALE) etc. |
| Model internals | Some parts of model inner structure. | Weights in linear models or the learned tree structure of the decision trees etc. |
| Data point | Some crucial points that can decide the prediction results. | Counterfactual examples, adversarial examples and some features highly contributed to the prediction results etc. |
| Intrinsically interpretable model | ML-based models that have interpretable structures. | DT, DR, KNN, and other linear models which have interpretable structures. |

**Attribute4: Model-specific or model-agnostic?** Model-specific attribute is limited to specific model types, which include a linear model, random forest, decision trees, neuro networks etc. Model-agnostic attribute can be found on any post-hoc method without opening a black-box model. These methods usually interpret the prediction results by evaluating the relevance between input features and outcomes, and completely skip the model inner structures.

### 2.3 Evaluation indicators towards interpretability results

When evaluating the model interpretability, we need to find evaluation indicators. The authors in [46] used properties of explanations as interpretability indicators to evaluate the model interpretability effectively. These indicators can be used to judge the performance of the explanation process. However, it is not clear how to quantify these indicators so as to be used in the specific interpretability work. Nonetheless, we can still divide the evaluation indicators into two categories, the first is for evaluating an interpretation method and the second is for evaluating the results generated from these methods in the specific experiments.

The first category has the following four evaluation indicators [46]:

1) **Expressive power**. It describes how well the interpretation method can interpret the prediction results. For instance, building the intrinsically interpretable models such as DT and DR can have a high expressive power to interpret a model or perhaps open a black-box.

2) **Translucency**. It describes how much an interpretation method can rely on looking into the inner structure of one ML-based model. For instance, the linear regression models are highly transparent thus we merely require a detailed analysis on the inner structure of these models to get an ideal interpretability. High translucency will lead to the method having the capability of generating more interpretation, whereas low translucency will lead to an interpretability method more portable.

3) **Portability**. It describes the range of ML-based models with which the interpretation methods can be applied. Methods using a surrogate model might have the highest portability, whereas methods that only work specific models will have lower portability.

4) **Algorithm complexity**. It describes the computational complexity of an interpretation method. As is known, the computation bottleneck always becomes a crucial problem when processing a huge dataset, not alone with the interpretation work on a great number of input features. For instance, activation maximization can print out layers outputs to provide interpretability but the algorithm complexity may become heavy.

The second category has the following seven evaluation indicators [46] of the specific experiment interpretability results:

1) **Accuracy**. It describes how well the interpretation predicts unseen data. High accuracy is crucial for prediction generated from ML-based models. Nonetheless, if we only focus on interpretability of model intrinsic logic, fidelity is the only important part without accuracy coming into consideration.

2) **Fidelity**. It describes how well the interpretation approximates the prediction of black box models. Accuracy and fidelity are closely related and if the black-box model has high accuracy and the interpretation has high fidelity, the interpretation towards the model also gets high accuracy.

3) **Consistency**. It describes how much an interpretation result differs between models trained on the same dataset that produce similar predictions.

4) **Certainty**. It describes whether an interpretation result reflects the certainty of the ML-based models. We need to get trust by these interpretation results so as to have full confidence to use the related interpretation method in another case.

5) **Degree of importance**. It describes how well our interpretation reflects the importance of features. For instance, if a decision rule is generated as an explanation for an individual prediction, we need to evaluate the degree of importance of the rule condition.

6) **Novelty**. It describes whether our interpretation reflects a data instance to be interpreted comes from "a new region far removed from the distribution of training data". This concept is closely related to the concept of certainty.

7) **Representativeness**. It describes how many instances an interpretation may cover. In some cases, especially in malware detection, we cannot always get a good tradeoff between interpreting the entire model effectively and interpreting the outcomes effectively. Therefore, we need to highly take into consideration representativeness as a crucial indicator for interpretability.

## 2.4 Classification of the existing common interpretation methods of ML-based models

The authors in [46] categorized common ML-model interpretation methods into model-specific methods, model-agnostic methods, example-based methods, and neural network interpretation methods. This section first introduces these four types in Section 2.4.1-2.4.4. Then, these four types are compared via several indicators in order to give researchers a good insight for selecting the appropriate methods towards different cases.

### 2.4.1 Model-specific methods

Using intrinsically interpretable models is called model-specific methodology, and this methodology can only focus on one specific model.

The major interpretable models are linear regression models, logistic regression models, GLM, GAM, DT, DR, RuleFit, and other interpretable models [46]. All of these models can be interpreted by using intrinsic model interpretation methods to give global interpretation. A different model has a different insight in its own interpretation formats. We detail the major interpretable models as follows.

For linear regression models [47][62][63], the advantage is that they are simple to be implemented and interpreted with the output coefficients and the regression algorithms to predict the samples are simple when confirmed to be used. In addition, linear regression models can avoid overfitting by using dimensionality reduction, regularization and cross-validation. However, the limitations are that the outliers have huge effects on regression and boundaries, and models assume independence between attributes. Furthermore, they cannot represent the feature relationships properly.

For logistic regression models [62][63], they are easy to implement, interpret, and train. Besides, they make no assumptions about distributions of classes in feature spaces. The limitations are that they assume the linearity between dependent variables and independent variables. In addition, they can merely be used for discrete functions.

For the generalized linear model (GLM) [86], they keep the weighted sum of the features but allow non-Gaussian outcome distributions, and they connect the expected mean of this distribution and the weighted sum through a possibly non-linear function. The limitations are that the interpretation of a weight can be unintuitive and the predictive performance cannot be ideal. Any link function that is not identity function complicates the interpretation. In addition, they rely heavily on assumptions about the data generating process.

For generalized additive model (GAM) [87], the strength is that they simply allow the generalized linear model to learn nonlinear relationships by relaxing the restriction. The relationship must be a simple weighted sum and assuming the outcome can be modeled by a sum of arbitrary functions of each feature. However, the limitations are that they have less interpretation due to the modifications on models, interactions and nonlinear feature effects. In addition, they rely heavily on assumptions about the data generating process.

For decision trees (DT) [88][89][90], the advantage is that they can solve the problem of linear and logistic regression. In addition, they can capture interactions between features and understand the distinct groups. Lastly, trees have a natural visualization and create good interpretability. However, the limitations of DTs are obvious. They fail to deal with linear relationships and lack smoothness. Besides, trees are also quite unstable. The number of terminal nodes increases quickly with depth so as to make interpretation difficult.

For decision rules (DR) [49][91][92], the advantage is that IF-THEN rules are easy to interpret. Decision rules can be as expressive as decision trees while being more compact. The prediction with IF-THEN rules is fast. Besides, DRs are robust and select only the relevant features. Simple rules like One-R can be used as baseline for more complicated algorithms. The limitations are that they completely neglect regression and the features have to be often categorical. In addition, DRs are weak in describing linear relationships between features and outputs.

For Rulefit techniques [93][94], the advantage is that they can integrate the feature interactions. In addition, they can become simple as linear models by learning a sparse linear model with the original features and also a number of new features that are decision rules. The limitations are that sometimes they create many rules that get a non-zero weight in the lasso model and become less interpretable with the increasing number of features. Besides, the weight interpretation is still unintuitive.

### 2.4.2 Model-agnostic methods

A model-agnostic method means separating the interpretation from the ML-based models, and thus has an advantage of interpretation flexibility than when using a model-specific method. In another word, we can use these methods to interpret any type of models. These methods include partial dependence plot (PDP), individual conditional expectation (ICE), accumulated local effects (ALE) plot, feature interaction, permutation feature importance, global surrogate, local surrogate (LIME), scoped rules (Anchors), LEMNA, Shapley Values, SHAP (Shapley Additive interpretation). We detail these methods as follows.

For partial dependence plot (PDP) [95][96][97], the advantage is that the computation is intuitive. In some cases, the interpretation is clear if the features are independent. PDPs are easy to implement and also have a casual interpretation for the calculation. The limitations are that the realistic maximum number of features in PD function is two. Some PDPs cannot show the feature distribution. The assumption of independence is the biggest issue for interpreting. Lastly, heterogeneous effects might be hidden.

For individual conditional expectation (ICE) [98][99], the advantage is that they are even more intuitive to understand than PDP, and some parts of the plots can represent the predictions for one instance if we vary the feature of interest. ICE curves can uncover heterogeneous relationships. The limitations are that the curves can only display one feature meaningfully and heavily depend on the independence of input features. Besides, if too many ICE curves are drawn, the plots can become overcrowded. ICE curves cannot be easy to see the average.

For accumulated local effect plot (ALE) [100], the strength is that ALE plots are unbiased that can be used when features are correlated and faster to compute. In addition, they can be interpreted by centered at zero and the 2D ALE plot only shows the interaction. The limitations are that ALE plots can become a bit shaky due to a great number of intervals and no perfect solution for setting the number of intervals. In addition, ALE plots are not accompanied by ICE curves, and the implementation of ALE plots is much more complex. Second-order ALE estimates have a varying stability across the feature space, which is not visualized in any way and a bit annoying to interpret. Lastly, unbiased structure cannot ensure the proper interpretation when features are strongly correlated.

For feature interaction (FI) [101][102][103], the strength is that the interaction H-statistic has an underlying theory through the PD decomposition and has meaningful interpretation. The statistic is dimensionless and detects all kinds of interactions regardless of forms, and analyze arbitrary higher interactions such as the interaction strength between more than 3 features. Nonetheless, the limitations are that H-statistic takes a long time to compute estimating marginal distributions which have a certain variance. In addition, this method cannot be available in a model-agnostic version so far and cannot interpret well in strong correlated features.

For permutation feature importance (PFI) [104][105][106], the advantage is that ideal interpretation, highly compressed and global insight can be given. They are comparable across different problems and automatically take into account all interactions with other features. In addition, PFI does not require model retraining. However, the limitations are that it is very unclear whether we should use training or testing data to compute the feature importance and the method is linked to the error of the model. Besides, we need access to true outcomes whereas the results may vary greatly. The method can be interrupted by correlated features and adding a correlated feature can decrease the importance of the associated feature as a tricky thing.

For global surrogate [107][108][109], the advantage is that surrogate models are flexible, intuitive and straightforward for the approach. In addition, with the R-squared measure, we can easily measure how good our surrogate models are in approximating the black-box predictions. Nevertheless, the limitations are that the surrogate model never generate the real outcome, and the interpretation of the model is not clear about the best cut-off for R-squared. In addition, a surrogate model can be very close to a black box for one subset of the dataset but widely divergent for another subset.

For local surrogate (LIME) [110][111][48][58], the strength is that we can use this for local interpretation on any given model. When using lasso or short trees, the resulting interpretations are short, selective and possibly contrastive. LIME can have fidelity measure and be easy to use. The local interpretation outcomes can use other interpretable features in addition to the original model features. The limitations are that it is not easy to find a proper kernel setting and a correct definition of the neighborhood is an unsolved problem when using LIME with tabular data. It is not easy to trust the LIME outcome and further exploration needs to be improved.

For newly-developed architecture (LEMNA) [20], compared with LIME, the strength is that it can help to set up a steadier trust on interpretation outcome than LIME because using LEMNA can approximate a local decision boundary correctly. In general, LEMNA is based on two insights. First, a mixture regression model can approximate both linear and non-linear decision boundaries based on enough data points. Second, 'fused lasso' is a penalty term commonly used for capturing feature dependency. By adding fused lasso to the learning process, the mixture regression model can take features as a group and thus capture the dependency between adjacent features. In this way, LEMNA can produce a high-fidelity interpretation result by simultaneously preserving the local non-linearity and feature dependency of DL models. However, the limitations are that we need to modify the input features ourselves, and it lacks the absolute interpretation effect when features become strongly correlated via the verification test.

For scope rules (Anchors) [112], the advantage is that it is easier to understand, as rules are easy to interpret than LIME. Furthermore, anchors can work when model predictions are nonlinear or complex. In addition, the algorithm is model-agnostic and highly efficient as it can be parallel used. However, the limitations are that the algorithm suffers from a highly configurable and impactful setup, and many scenarios require discretization. Constructing anchors require many calls to the ML model. Lastly, the notion of coverage is undefined in some fields.

For Shapley values [113][114][115][59], the obvious strength is that the difference between the prediction and the average prediction is fairly distributed among the feature values of the instance. The Shapley value allows contrastive explanations and is the only explanation method with a solid theory. It is mind-blowing to explain a prediction as a game played by the feature values. Nevertheless, the limitations are that the Shapley value requires a lot of computing time. The Shapley values can be misinterpreted and explanations created with the Shapley values method always use all the features. In addition, it cannot have a prediction model like LIME. We need access to the data, whereas it suffers from the inclusion of unrealistic data instance.

For Shapley addictive interpretation (SHAP) [116][117], all the advantages of Shapley values are added to this method, and the SHAP has a solid theoretical foundation in game theory. The prediction is fairly distributed among the feature values. We get contrastive explanations comparing the prediction with the average prediction. SHAP, connecting LIME and Shapley values, has a fast implementation for tree-based model. Less computation overhead makes it possible to compute many Shapley values needed for the global model interpretations. However, the limitations are that the kernel of SHAP is slow and ignores feature dependence, and tree versions of SHAP can produce unintuitive feature attributions.

### 2.4.3 Example-based methods

Example-based methods can interpret the models by selecting particular instances of the dataset to explain both the behaviors of ML-based models and the underlying data distribution. There are several methods based on examples which are counterfactual interpretation, adversarial interpretation, prototypes and criticisms, and influential instances, etc. We detail these methods as follows.

Counterfactual interpretation [118][119], the outcome of counterfactual explanation is effective for local interpretation. The counterfactual method does not require access to the data or the model, and it works with systems that do not use ML. In addition, the method is relatively easy to implement. The limitations are that, for each instance you will usually find multiple counterfactual explanations (Rashomon effect), there is no guarantee that a given tolerance is found, and the method cannot handle categorical features with many different levels well.

For adversarial interpretation [120][121][122], these adversarial examples make models vulnerable be attacked, which gives models local interpretation. However, the limitation is that the method to generate the adversarial examples require further exploration.

For prototypes and criticisms [123][124], the advantage is that the participants performed best when the sets showed prototypes and criticisms instead of random images of a class. Besides, the number of prototypes and criticisms can be decided by ourselves. Among the criticism methods, MMD-critic works with density estimates of the data, any type of data and any type of machine learning model. The algorithm of MMD-critic method [136] is easy to implement. Finding criticisms are independent in the selection process of the prototypes. Nevertheless, the limitations are obvious. While, mathematically, prototypes and criticisms are defined differently, their distinction is based on a cut-off value which is the number of prototypes. Besides, these methods cannot work well if you do not choose prototypes and criticisms. To select a kernel and its scaling parameters is a very important issue. The prototypes and criticisms-based methods take all the features as input, disregarding some features that might not be relevant for predicting the outcome of interest. There are some codes available, but it is not yet implemented as nicely packaged and documented software.

For influential instances [125], the advantage is that it requires highly focus on influential instances emphasized by training process, and becomes one of the best debugging tools for ML-based models. Deletion diagnostics are model-agnostic, and we can use them to compare different models. Influence functions via derivatives can also be used to create adversarial training data. The approach is generalizable to any question of the specific form. When it comes to the limitations, it is expensive to calculate. Influence functions are a good alternative to delete diagnostics, but only for models with differentiable parameters. The functions are only approximate, because the approach forms a quadratic expansion around the parameters. There is no clear cut-off of the influence measure at which we call an instance influential or non-influential. The influence measures only take into account the deletion of individual instances and not the deletion of several instances at once.

### 2.4.4 Neuro network interpretation methods

To make a prediction by a neuro network, the data input is processed through many layers with the learned weights and non-linear transformations. We cannot follow the exact mapping from data input to outcome generated by DNN or any other neuro network. Although we can use some model-agnostic methods to interpret neuro network, it cannot be interpreted well if we do not develop these methods to a specific version suitable with neuro networks.

In general, there are two reasons for classifying some methods into neuro network interpretation methods type. Firstly, neuro networks learn features in their hidden layers and we need specific tools to uncover them. Secondly, the gradient can be utilized to implement interpretation methods which are more effective than the model-agnostic methods. Finally, feature visualization, example-based interpretation and learned features are three methods used for neuro network. We detail these methods as follows.

For feature visualization [126][127], the advantage is that it can give a unique insight into the working of neuro networks, and that network dissection allows us to au-tomatically link units to concepts and detect concepts beyond the classes in the classification task. In addition, it is a great tool to communicate in a non-technical way how neural networks work and can be combined with feature attribution methods. Nonetheless, the limitations are that many feature visualization images are not interpretable, and there are too many units to focus on. For network dissection, we need datasets that are labeled on the pixel level with concepts which are hard to collect.

For gradient-based interpretation [128][129][130][131], the strength is that it can be utilized to implement interpretation methods that are more computationally efficient than model-agnostic methods. However, the limitations are that the gradient can vanish when layers are too many and cannot solve the problem of correlation between input features, such as Saliency Masking (SM) cannot consider the relationships between the input features from a strong semantic sequence.

For learned features method [132][133], the advantage is that it can use features learned from layers and networks, and it can be applied with comprehensive approaches. The limitations are that the formats of data inputs need modified and computational cost is variable. Besides, mechanism layer embedded onto the model needs further exploration like LEMNA.

### 2.4.5 Comparison and evaluation on the common interpretation methods in groups

This section describes the evaluation indicators for interpretability, which are used for the detailed comparison of the methods discussed in Section 2.4.1-2.4.4. The explanations of these indicators are as follows.

Firstly, the visualization is intuitive for researchers to better understand the interpretation. Secondly, whether our method can consider the correlated features from the input data will directly restrict the scenarios our method can be used for. For instance, if we want to interpret the NLP problems and prediction results, then the input features would be commonly correlated whereas some problems are independent from the feature relevance. Thirdly, we consider the expressive power, translucency, portability and algorithmic complexity as our evaluation indicators towards comparing interpretation methods, and we set three levels to evaluate these indicators, namely low, medium, and high. Note that we cannot quantify these factors by weights but basically evaluate them by three levels. These indicators are directly described in the first part of Section 2.3.

**Table** 3 depicts the comparison of the existing interpretation methods from these indicators. We can observe that different group methods have different interpretability performance. In general, model-specific methods have high translucency and low portability. Model-agnostic and example-based methods have high portability and low translucency but expressive power often becomes stronger. In particular, interpretability will become more comprehensive when outcome is global and the algorithm is less complicated. Besides, the correlated features always interrupt the final explanation, thus only a small number of all methods solve this restriction such as LEMNA, rule-based approaches, DT, ALE, feature interactions, and learned features in neuro network-based approaches.

As mentioned above, we can choose different methods for different scenarios according to the evaluation results. For instance, if we want to process a sequential problem, then the approach must cover the consideration of correlated features which thus narrows the search range. Likewise, if we want to process a problem based on the portable interpretation methods, then we should select the methods such as example-based methods, permutation feature importance, and global surrogate and so on. Note that not all model-agnostic methods are portable such as feature interaction method because it has not been yet tested via experiments. Besides, if we consider about the algorithmic complexity when using these methods, then the neuro network interpretation methods would not be a good option, instead we should choose methods like some intrinsic models, PDP, permutation feature importance, SHAP, and example-based methods etc. Lastly, if we want to consider both the low complexity of algorithm and the strong expressive power for instance, then there would be DT, DR and RuleFit for selection. In such a way, we can choose a relatively better method for the specific interpretation in our own situation.

In general, although we can evaluate and compare these methods in groups and find out their advantages and limitations in different scenarios, there is no description about one absolutely great method in each group. Hence, we should select different methods in the different scenario. Basically, the main issues for selecting the ideal interpretation methods come into the following several points:

1) When features become more polymorphic, our model criticisms would break down and no longer suitable for the designed prototype. This is highly due to the reason that most work only gives local interpretation, as is illustrated in the table.

2) Computational cost always reminds us to design a lightweight framework that can still be applied for a huge dataset. The evaluation indicators of interpretation described in Section 2.3 always remind us to give interpretation models high portability and low algorithm complexity.

We note that when the interpretation is model-agnostic, it often gives local and unsteady interpretation. Therefore, model-specific approaches can improve the interpretation fidelity and we can choose them in most cases if allowed.

## 3  The interpretability of malware detection

In Section 2, we introduce the concept of interpretability, the attributes of interpretation methods, the evaluation indicators, and the related work in the common field. Based on them, we then step into the interpretability in the specific malware detection field because the interpretability work can be more valuable in specific fields.

This section includes the following parts:

1) The importance of ML-based malware detectors: why we should interpret these detectors in many cases in Section 3.1.

2) Current limitations and challenges faced by interpreting these detectors in Section 3.2.

3) The solutions to mitigate the unresolved limitations faced by interpreting the detection: the traditional optimized approaches and the state-of-the-art approaches towards interpreting the malware detectors in Section 3.3.

4) A new taxonomy of interpreting malware detection approaches and detailed evaluation towards the related work about these state-of-the-art interpretation approaches in malware detection.

5) Some discussions on methods which cannot be categorized into the aforementioned taxonomy and open research issues towards malware detection interpretability.

**Table 3 The comparison and evaluation on the common interpretation methods in groups: note that interpretability evaluation factors in groups are visualization, correlated features covered, expressive power, translucency, portability and algorithmic complexity. Besides, H represents high degree, M represents medium degree, and L represents low degree**

| Group | Methods | Ref. | Interpretability evaluation | | | | | |
|---|---|---|---|---|---|---|---|---|
| | | | Visualization | Correlated features solved | Expressive power | Translucency | Portability | Algorithmic complexity |
| Model-specific methods | Linear regression models | [47][62][63] | × | × | M | H | L | L |
| | Logistic regression models | [62][63] | × | × | M | H | L | L |
| | GLM | [86] | × | × | M | H | L | L |
| | GAM | [87] | × | × | M | H | L | L |
| | Decision Trees (DT) | [88][89][90] | √ | √ | H (short depth); L (long depth) | H | L | L |
| | Decision rules (DR) | [91][92][49] | × | √ | H; L (linear relationships) | H | L | L |
| | RuleFit | [93][94] | × | √ | H; L/M (features increasing) | H | L | L |
| Model-agnostic methods | PDP | [95][96][97] | √ | × | M | L | M | L |
| | ICE | [98][99] | √ | × | M (better than PDP) | L | M | L |
| | ALE | [100] | √ | √; × (when features are strongly correlated) | M (better than ICE, PDP) | L | M (a bit difficult to be applied for complex structure) | M (too many intervals) |
| | Feature interaction | [101][102][103] | √ | √; × (when features are strongly correlated) | H | H | L | H |
| | Permutation feature importance | [104][105][106] | √ | × | M (linked to errors interpretation) | L | H | L |
| | Global surrogate | [107][108][109] | × | × | H (global interpretation) | L | H | H/M/L (depends on the surrogate) |
| | LIME | [110][111][48] | √ | × | M (local interpretation) | M | M | M |
| | LEMNA | [20] | √ | √ | H (by rules) | M | H | M |
| | Scoped rules (Anchors) | [112] | √ | × | H | M | H | H/M/L (highly variable due to many calls to the ML models) |
| | Shapley Values | [113][114][11559] | √ | × | H | L | H | H |
| | SHAP (Shapley additive interpretation) | [116][117] | √ | × | M | L | H | L |
| Example-based methods | Counterfactual interpretation | [118][119] | √/× | × | H | L | H | L |
| | Adversarial interpretation | [120][121][122] | × | × | H | L | H | L |
| | prototype Criticisms | [123][124] | × | × | H | L | H | L |
| | Influential instances | [125] | × | × | H | L | H | L |
| Neuro network interpretation methods | Feature visualization | [126][127] | √ | × | H/L | L | H | H |
| | Gradient-based interpretation | [128][129][130][131] | √ | × | M | L | M | H |
| | Learned features | [132][133] | √/× | √/× | H/M | M | H | H |

### 3.1 Importance of ML-based malware detector interpretability

We know that ML can help save human labor extensively by mainly contributing on these three reasons: 1) Automatically learn common patterns of malware from the feature space rather than manually craft signatures based on traditional feature extraction using dynamic and static ways (features like API calls and API trace sequences etc.). 2) Do not rely on simple signatures but comprehensive combination of patterns identified from space. 3) Do not require human experts to manually examine samples [1].

With a wide variety of ML-based malware detection approaches available, researchers would find it challenging to select the best ones to improve the results of malware detection. Among classification towards malware samples, not only do we need to select the better features to be fed up into classifiers, but also the accuracy and interpretability may affect the performance of ML-based malware detection. Over the main structure of the traditional feature selection and ML-based classification, malware family clustering is the critical part to defining the final results in no doubt. **Fig. 1** shows the traditional feature extraction and classification based on ML in expert field.

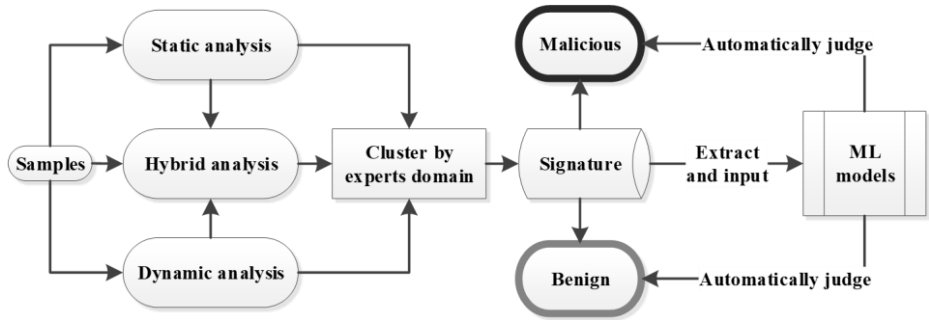

**Fig. 1 Traditional structure of feature selection and clustering whose features are then fed up into ML**

According to [8], accuracy is a measure of how well our classification algorithms perform in practice, while interpretability refers to the capability to explain how a single decision is made. In addition, the interpretability work can only be worth exploring under the circumstances that the prediction accuracy is very high. In malware detection field, if we cannot interpret the ideal prediction results, we will be limited in our deployment of detectors due to the fact that when using black-box models, a wrong result will have catastrophic consequences.

Hence, we conclude that in order to widely deploy the models in cyber environment, the interpretability is far beyond doubt a key point to optimize the current malware detector systems based on ML as well as enhance security trust to the models.

### 3.2 Limitations and challenges faced by ML-based malware detection interpretability

Malware detection models have two major limitations, one comes from model interpretability, the other comes from their limited processing capability on long malware assembly code. To solve these two issues, DL-based and some improved ML-based detection models can mitigate the processing bottleneck, whereas these models always have black-box attributes. Hence, there must be a trade-off between accuracy and interpretability in malware detectors.

Specifically, based on the above issues, different DL-based detectors may have different limitations to some degree. The instances are as follows.

CNN and RNN are two major methods used for DL-based malware detection, but CNN-based approaches drop their accuracy when the length of sequence of malware opcodes and samples called by interdependent library becomes variable. This problem requires long preprocessing time and large-scale training data even that we use CUDA GPU for accelerating the speed. RNN approaches can be used for mitigating the former limitation in deep neural networks using multi-layer neural network, but due to the problem of gradient vanishing when layers become too many, the traditional RNN approaches can only generate the relatively short-length features.

To address the limitation faced by RNN, LSTM network [4] can be considered. The related varied versions of this model such as in [5][6][7] have been developed to further alleviate the limitation.

Moreover, another possible strategy to solve this limitation is trying to separate the assembly code of an executable into short fragments, such as using n-grams, which has been explored and performed well in many experiments [17]. However, this strategy has two main drawbacks. One is that the overall semantic meaning of malware sequences may be inevitably lost. The other is that matching assembly functions of unknown executables with functions of known malware samples is always based on little similarity, thus a result derived from n-grams always fails to detect the obfuscated malware variants.

In general, when developing a new interpretable DL model, we need to consider the following challenges:

1) Although approaches based on DL models generally achieve high malware classification accuracy and do not need to use manually preprocessed features generated from experts' experience, most DL-based models will cost high memory usage due to the long length of malware feature sequences.

2) When generating the rule-based interpretation, most results are local and cannot be used for putting a trust on deploying detectors in the wild for their low fidelity.

3) The existing interpretation methods have certain weaknesses in specific situation more or less, and a new better method is still being explored by many researchers.

4) There is no certain quantitative indicators system for measuring interpretability and lots of work still stay at the level of qualitative description.

5) Commercial-level products have never appeared.

### 3.3 Solutions to mitigate the unresolved issues towards malware detection interpretability

In this section, we follow the unresolved problems described in the last section and point out the basic solution of malware detection interpretability challenges. In general, most detectors models are black-box, and the common approach workflow for interpreting these detectors is illustrated in **Fig. 2.**

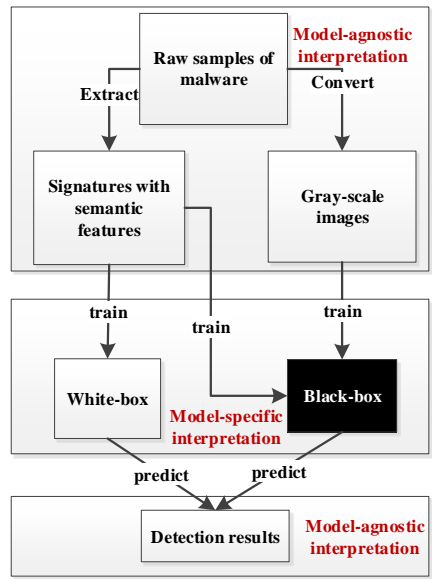

**Fig. 2 The workflow of interpreting ML-based malware detectors**

As **Fig. 2** shows to us, we can interpret many phases in a model to get trust on the final results. Black-box detection model like DL-based such as CNN and RNN and like traditional ML-based such as SVM and nonlinear classifiers, fed up by features becomes difficult to interpret but easy to extract from raw datasets and highly-processed features. In contrast, white-box detection model like Decision Tree and linear classifiers, fed up by features can be fully understood and highly processed by expert knowledge.

Section 3.3.1 gives a brief overview of traditional malware detection methods based on signature and some optimized versions of malware detection methods such as API-based detection architectures aiming at enhancing the interpretability of the ML-based malware detection models. In addition, we evaluate these optimized versions through several indicators.

Section 3.3.2 gives the state-of-the-art interpretation framework towards ML-based malware detectors. Note that Section 3.3.1 mainly involves the right part of **Fig. 2**, and Section 3.3.2 mainly involves the left part of **Fig. 2**.

### 3.3.1 The optimization progress logic of interpretation methods of traditional malware detection models

The signature-based approaches to analyzing malware samples have turned out to be static analysis, dynamic analysis, and hybrid analysis combined with the former two methods to extract some multimodal features. On the one hand, static analysis extracts features derived from datasets without involving their execution and just analyze the code written inside the malware samples. The static analysis can be quick and cover the full file but suffer from obfuscation techniques. Obfuscation is used to hide malicious behaviors and then crack the static analysis system. On the other hand, dynamic analysis derives features from executing samples in the virtualized environment using sandbox like Cuckoo. These features include most sensitive API call traces and behaviors such as attacking Windows Registry, network, files, PDFs. The dynamic analysis cannot be cracked by the obfuscating samples but waste lots of running time. Moreover, this kind of analysis can be discovered by the anti-debugging modules and fails to detect the malware including these modules. Given the pros and cons of static and dynamic analysis, hybrid analysis combines the advantages of both analysis methods by extracting multimodal features. The general feature taxonomy structure is shown as **Fig. 3**.

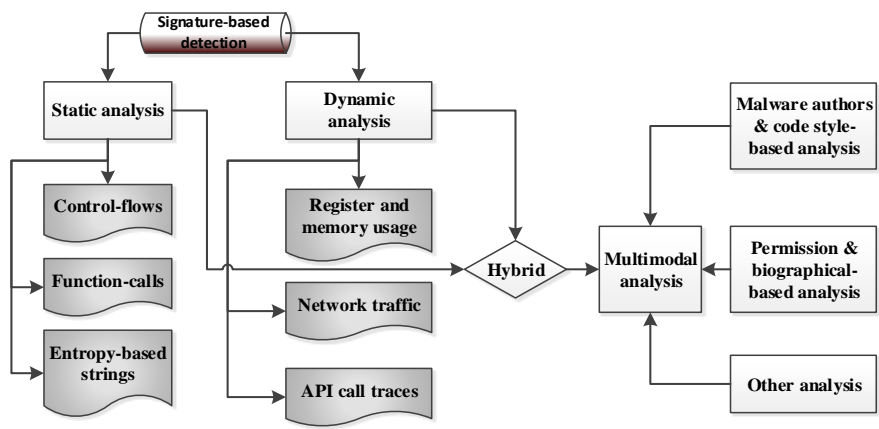

**Fig. 3 Main structure of signature-based feature selection**

As is illustrated in **Fig. 3**, signature-based detection can have many types of features to extract. In static analysis, control-flows, function-calls, and entropy-based strings are three main types. In dynamic analysis, register and memory usage, network traffic, and API call traces are three main types. To hybrid them, researchers use multimodal analysis to enhance the detection accuracy. These manually extracted features can be understood by the final classification result to some degree, and saved into the signature database. Moreover, ML-based detection can use these preprocessed features to predict the unknown variants and get interpretation to some degree. However, in this way, it is of no necessity to using DL-based models because KNN and other linear models may have a better result.

Therefore, current research work still use the traditional signature-based malware detection approaches [32][33][34][35] and get high accuracy. Among the selected features, API call information derived from hybrid analysis turns out to be the most interpretable semantic features due to the interpretability of semantics. In general, researchers can extract different API information and combinate them in order to plot a directed graph as the input feature of classifiers. These digraphs are generated from the dependence of the arguments and return values. The related structure based on API call combination and detection is shown as **Fig. 4.**

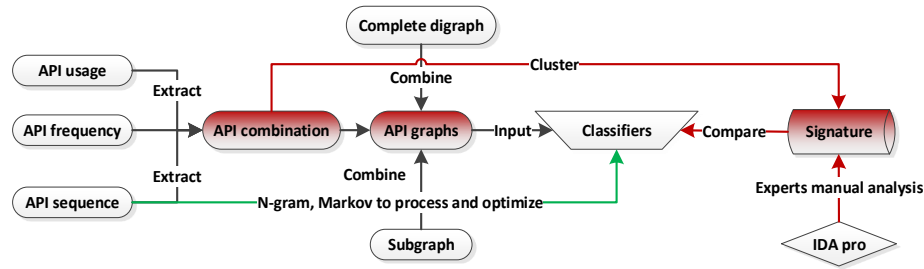

**Fig. 4 The studying structure based on API information combination and detection: note that the red line refers to clustering the API combination into signature databases, and the green line refers to converting API sequences into n-gram, Markov graphs as the input data of classifiers**

To mitigate the issue of interpretability, we should improve the traditional method based on hybrid analysis to generate the API call information on a small-scaled dataset. **Table 4** mainly shows the current work for these optimized traditional approaches using API call information. Here we highlight the usage of graph generalization approach based on extracting API calls using the relevance between each two calls parameters as the directed graph edges. In addition, by observing and comparing different work, we consider the impact of the integrity of the graph on the final detection results.

From **Table 4** we can know that API information can have some optimization processes (such as transferred into graphs and features processed) to generate features fed up into improved traditional analysis detector based on ML. These API-based approaches are highly interpretable due to that the formats of the features are semantic rules and API call graphs behave even more interpretable due to the high logic existing in the malware behaviors. The data mining towards these graphs and their completeness can give comprehensive API-based semantic interpretation due to the fact that these rules are generated from behavior analysis. Since rules are understood by human beings, the API-based semantics can give dependable prediction and the interpretation results are mostly generated from signatures. For instance, Cuckoo analyzes a malware sample by matching the signatures and these signatures which mainly include API calls interpret the detection result.

From this angle, we cannot completely abandon this traditional approach because it can get rid of manual expert workflow to some degree and get an ideal trade-off between interpretability and accuracy. Furthermore, with the API call digraphs and combination information, we can interpret the results by analyzing these features, which involves data mining approaches.

**Table 4. Traditional optimized detector approaches based on API information: note that we cannot fully assure its prediction accuracy when applied onto another dataset.**

| Ref. | Platform | Uncomplete graph | Complete graph | Prediction Accuracy | Features | ML/Dedicated methods | Datasets |
|------|----------|------------------|----------------|---------------------|----------|----------------------|----------|
| [32] | Windows | × | × | 98.64% | API call sequences&&behavior chains | LSTM (depth network) | n/a |
| [35] | Android | √ | × | 98.98% | API combination: API frequency, API usage, API sequence detection | DT (C4.5) | 10956 benign samples in 2014 from Play Drone, 4,000 new apps in 2018 from Play Store, and 28,848 malicious samples from Virus-Share |
| [33] | Windows | √ | × | 96% (labeled); 91% (achieved) | API directed graph (uncompleted), API calls | RF 90.3% (highest TP for Drebin and OB); RF 95.7% (highest TP for Marvin); and others like SVM, K-NN have less TP rate which don't show here | 53,361 benign and 54,324 malware samples from the Windows system |
| [34] | Android (Linux, varied Linux) | √ | × | 99.8% | Event group (higher level than API information), which later turns into function clusters and API calls vectors | NNs (SVM NB DT); EveDroid; MaMaDroid | 43,262 benign and 20,431 malicious samples from Marvin (2015), Drebin (2014), OldBenign, NewBenign, ObData (2015), PackedApps (2018) |
| [38] | Android | × | √ | 98.86% | API call graph | Deep neural network and maximized-effective performance training network for different embedding | Benchmark:AMGP(2011)&Drebin(2012) In-the-wild: Malware (2014) Benign (2014) |
| [39] | Android | × | √ | 91.42% | API call graph | Deep neural network and maximized-effective performance training network for different embedding | AMD (2017) AndroZoo (2017) Drebin Malware Collection (2014) ISCX Android Botnet (2015) |
| [36] | Android (Linux extended) | n/a | n/a | 97.89% | Network traffic | ML algorithm C4.5 | 10010 benign APPs from AndroZoo && 10683 malwares from AMD |
| [37] | Android | n/a | n/a | 99.2% (benchmark); 86.2% (in the wild) | Multi-level fingerprints feature hashing | The first method developed based on n-gram analysis and online classifiers (can be extended) | Drebin project (2014): 5560 real malware samples |

### 3.3.2 State-of-the-art interpretation logic for ML-based malware detection

As is depicted in the previous section, the API-based data mining can interpret the malware detection results to some degree. However, these interpretable methods rely heavily on manual work and lack raw data features because the raw samples may have some hidden features that cannot be extracted by manual work. Given this issue, DL models can skip the manual work directly and be proved to have high prediction in many experiments. Nonetheless, DL-based detectors are black-box models, and thus the interpretability work shifts to the aspect of opening these black-box detectors.

The related key approaches can step into two main aspects:

1) Understand the black-box classifiers workflow and develop a more interpretable model.

2) Interpret the input features from raw datasets, then weight their contributions to the final prediction results.

In general, it is difficult for us to get a highly interpretable detector with ideal prediction performance on raw datasets. Unless we do manual work to analyze the samples, a well-behaved malware detector with ideal prediction accuracy always performs on a black-box model, or the traditional interpretable ML-based model such as decision trees, rules, linear models with more layers or depths. When the structure of these models becomes more complicated, their interpretability may well become lower than a simple structure of DL with just several layers to train the datasets [82]. Hence, these detection models always require a comprehensive interpretation in the outcome or inner structure aspect.

### 3.4 State-of-the-art malware detector interpretability work

Based on the two aspects of opening the black-box models in Section 3.3.2, we then give a new taxonomy and evaluation towards the related state-of-the-art work compared to the existing taxonomy in Section 2.4.

The common model interpretability involves with five major problems according to [84]: model problem formulation, model explanation, outcome explanation, model inspection, and transparent box architecture. In the malware detection field, we believe we need a new taxonomy to classify all the related work by concerning the above five issues.

Hence, as is illustrated in **Fig. 5**, we classify the state-of-the-art malware detection approaches with interpretability modules into model interpretation aspect and outcome interpretation aspect. Note that model interpretation aspect focuses on issues like model problem formulation, model explanation, model inspection, and transparent box architecture. Outcome interpretation aspect focuses on outcome explanation issue.

The model interpretation aspect includes feature transfer, optimization and selection. Here we classify the specific techniques related to this methodology into six types, including image-based, rules-based, static-based, dynamic-based, n-gram-based and gradient-based methods. Moreover, this aspect also includes model reformulation whose major techniques are two (LIME and extended type && Deep-Taylor Decomposition), as illustrated in the right column of **Fig. 5.**

The outcome interpretation aspect includes exploring the data points like counterfactual examples and adversarial examples.

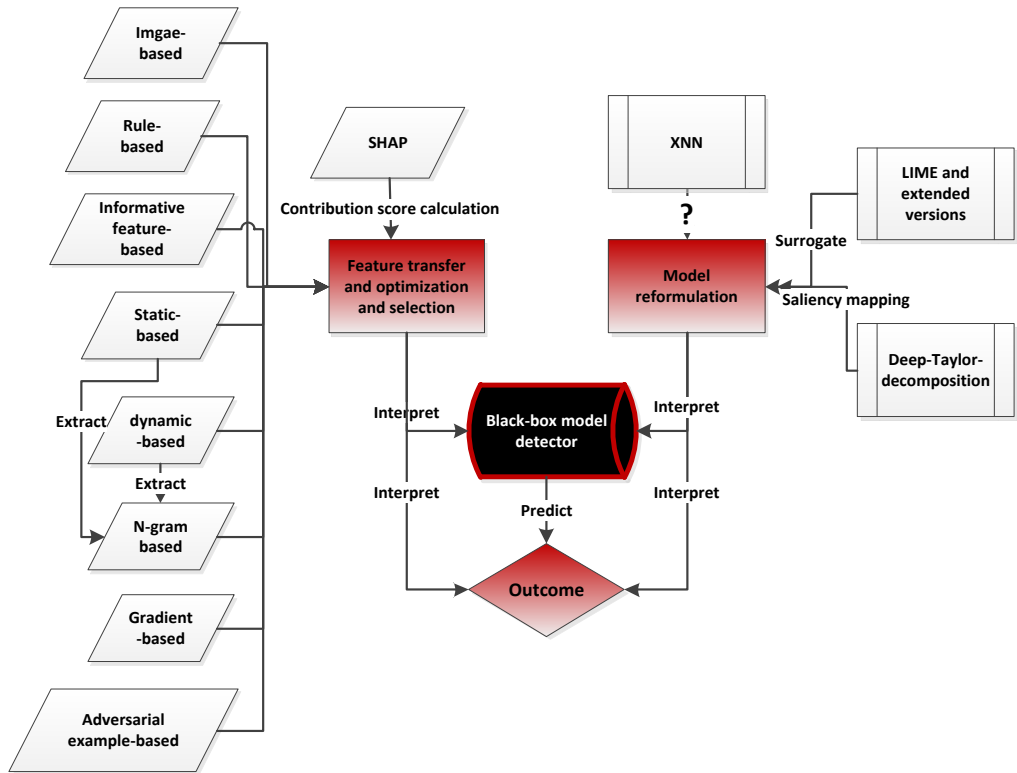

**Fig. 5 Taxonomy of malware detection model interpretation approaches**

Here, by reviewing the Reference, we find that the following five points mainly contribute to the current work.

First, visualization and using rules to interpret are crucial in a malware detector model. Second, the traditional malware detection based on signature mainly involves with static and dynamic analysis. The features generated from them can become signatures and give heuristic insight to help interpret the malware detectors. In addition, the informative features processed by novel strategies also enhance the interpretability. Third, gradient-based methods can give inspection into the inner structure of one ML-based model, which can help to interpret the prediction results generated from DL-based malware detectors. Next, using surrogate to interpret the black-box models and reformulating the model learning phase can give intrinsic interpretation. These methods include LIME, DTD (also known as saliency mapping methodology) and a potential surrogate used for interpretable detection called XNN. Then we have SHAP to calculate important feature contribution scores with negative and positive effects to select features for explanation. Lastly, adversarial example-based analysis can directly interpret the model prediction, which means that we can interpret the model outcomes via malware samples attacking technologies towards detectors. The adversarial samples are the data points in models we want to interpret. These samples might help us better interpret the models by their counterfactual features.

This section is organized as follows. Section 3.4.1 to Section 3.4.10 mainly describe the classified methods through the above taxonomy. Section 3.4.11 gives a new architecture requiring further exploration. Section 3.4.12 discusses the recent critical work towards interpreting malware detection and a newly written survey related to the future of this field. Section 3.4.13 evaluates and compares all the work in details through their strengths and limitations.

To our best knowledge, this is the first effort trying to provide such taxonomy.

### 3.4.1 Image-based

Image-based interpretation often refers to interpreting the grayscale images and some variants which can be visualized.

In [9], researchers applied DL architectures to detect the malware variants. They used deep transfer learning to get high accuracy and low FP rate compared to many classical ML algorithms. In their work, they utilized the local-interpretable model-agnostic approach and used the image-based DL classifier for interpretability, and as we know that DL models have a great performance on image processing. They selected a malware image from malware family Lloyda.AA2 and plotted 200 super-pixel representation, then they examined what aspects of the malware images were considered by the DL model to generate the prediction. In the final work, they used visual interpretation to get interpretation results, such as the red regions indicated the pixel regions that the model did not trust in producing the prediction.

In [10], compared to the traditional signature-based ML detection, researchers proposed a new byte-level malware classification called MDMC which converted malware binaries into Markov images and classified malware Markov images with DL. Their approach can be deployed in different platforms and has a better interpretation result than the DL based on grayscale-image representation. Furthermore, the approach can reduce the noise so as to fix the input data to a certain size. Their work based on the Markov model achieved a higher accuracy and better interpretability compared to GDMC based on grey-image malware formats.

### 3.4.2 Informative feature-based

Informative features refer to the features generated from novel methods and usually become patterns to enhance the interpretability. In this type of methods, the feature generation process is critical.

In [11], the authors proposed a novel feature selection method called informative variable identifier, capable of identifying the informative variables and relations. The approach can transform the input variable space distribution into a coefficient-feature space using linear classifiers or CME. After resampling techniques applied on these coefficients, the experiment results can outperform state-of-art algorithm with a good interpretability. Furthermore, they have extended previous definitions of interpretability in order to interpret the relations among features on redundant datasets. Main sections of the selection algorithm come into four steps: first transferring variable space to weight space, secondly generating relevant variables and redundancy among variables detection, then producing graphs and subgraphs as well as noisy variables, finally ranking all the informative features by descending order or importance.

In [41], the authors investigated feature obfuscation towards black-box malware detectors on Android platform. This work showed that a combination of features can simply dominate the overall detection results. The result has showed us that black-box analysis model should be interpreted by multimodal aspects. In addition, although HMM, SVM and DL models in [42] have given great promise on malware detection accuracy, the interpretability of these models are lacking in exploration.

For these informative features derived from samples, more comprehensive variables generated from using Hamming distance as the benchmark for the similarity-based malware detector classification is also applied in [73]. The similarity existing in API calls can generate the interpretability in their framework to some degree.

Opcodes-based models also contribute to this informative-feature based field which was further explored by Darabian et al. [76] in 2020 by the method of interpreting on different black-box models. Their work detected maximal frequent patterns (MFP) of opcode sequences to get interpretability of model prediction to some degree in order to help detect polymorphic IoT malware samples.

### 3.4.3 Rule-based

The rules are mostly understood by human beings, and using rules to interpret the decisions can be easily validated by expert knowledge.

In [12], the authors proposed a convolutional neural network (CNN) learning structure with added interpretability-oriented layers in the form of Fuzzy Logic-based rules. User can extract linguistic Fuzzy Logic-based rules from the DL structure and link this information to features derived from the preprocessing, thus to enhance the interpretability of the whole classification. Rule-based interpretability results can be ideal in the form of linguistic rules which link the classification decisions to the input features, but the raw malware data features must have some special attributes which require further exploration. In general, IF-THEN rules can help malware experts to interpret the prediction results.

### 3.4.4 Static-based

Features from static analysis alone can also be interpreted and these interpretation methods are often related to the static analysis knowledge and reverse engineering.

In [18], the authors proposed a novel interpretable malware detector using hierarchical transformer called I-MAD, which achieved ideal performance on static malware detection and understood assembly code at the basic block, function, and executable level. Furthermore, they also integrated their novel interpretable feed-forward neural network to provide interpretation for detection results by investigating the impact of features in response to the prediction. Their main contributions include:

1) Proposed a DL model to understand the full semantic meaning of the assembly code of malware executables.

2) Proposed two pre-training tasks to train bottom-level and middle-level components of H-Tran to interpret codes at basic block and function level.

3) Interpreted the feed-forward neural networks.

4) Can help malware analysts locate malicious payloads and patterns in malware samples.

Although static analysis can be cracked by malware packer obfuscation, a few researchers realize that benign samples can also have packers to crack the detectors. Authors in [72] explored this issue and further interpreted the prediction results by using static-based packers. Their work first tried to explore the packed Windows executables on ML-based malware classifiers which only use static analysis features. With respect to the malware packer detection, they can give more insight into interpreting the malware detectors.

### 3.4.5 Dynamic-based

The dynamic-based interpretation refers to the interpretation by using dynamic analysis and finding the correlation between dynamic behaviors and prediction outcomes.

In [28], the authors proposed a lightweight, accurate and interpretable automated malware detection system called NODENS. The limitations include two aspects. On the one hand, datasets used for training are on a small scale and only consist of 146 malware samples in total. On the other hand, their training samples were run in a virtualized environment which means that the system could not resist VMware malware. Nonetheless, this in-depth decision classifier has a good interpretability which offers weight to assessments drawn from analysis of raw data and makes it easier for users to analyze other malware samples by using easily understandable output formats. Finally, their investigation identified three critical malware distinction metrics, including binary values, output scores and differential between (Peak)WorkingSet64 and PrivateMemorySet64. Their dynamic-based approach basically remedies with [29][30][31].

The authors in [69] provided insight into the parameter tuning of ML algorithms for detecting the malware using API calls derived from dynamic runtime analysis based on Cuckoo. In this work, they got supervised ML assessed with 6434 benign and 8634 malware samples both to get high accuracy and great interpretation. Next, API calls and permissions were embedded to explore prediction results and forensic ability according to features of interpretability in [70].

In order to get high interpretability, we should also explore unsupervised data learning, which can help us get rid of the high accuracy performed only on the known labeled samples. Authors in [74] proposed providing a feature-based semi-supervised learning to detect malware from Android to achieve this goal in the dynamic analysis. Dynamic-based analysis sometimes involves permissions that malware accessed when running on a monitor. Due to this fact, using permissions can be also effective to predict, and authors in [75] have used this method to evaluate the robustness of textual descriptions for malware detection. In general, dynamic-based interpretation methods can cover permissions and API calls etc.

### 3.4.6 N-gram-based

The n-gram-based interpretation can use n-grams generated from hybrid analysis, and use heuristic learning or other data mining techniques to explore the interpretability. N-gram-based analysis is used to handle with the Natural Language Processing (NLP) issues, and there have been some related work about using it to solve malware detection, but they are faced with a problem of too long length of sample contents, such as strings, API call traces. In these cases, the value of n is always set up with 2 or 3.

In [25], the authors developed interpretability approaches in practice and evaluated the effectiveness of existing interpretability techniques in the malware detection using the features derived from n-gram analysis. Their work interpreted the logistic regression model, random forest, a single sample with random forest, neural network model, and a single sample of a neural network. Their work first used n-gram processing method to give interpretability. In [135], the authors provided an adversarial machine learning method on malware detection model based on opcode n-grams features. Their main contributions include generating adversarial examples to attack the detectors and use n-grams feature to interpret the model prediction results.

In general, n-gram-based interpretation can be improved by the static and dynamic analysis, and a more complicated manual analysis towards input features always refers to a more detailed interpretation.

### 3.4.7 Gradient-based

The gradient-based interpretation uses gradients in the neuro network layers and opens the black-box with deep layers.

In [64], the authors proposed a framework to generalize malware detectors using linear and explainable ML models to any black-box ML models. They leveraged a gradient-based approach to identify the most influential local features and highlight the global characteristics learned by the model to discriminate between benign and malware samples. They provided with mean relevance scores around the benchmark datasets and tried to explore surrogating model methods to impact the explanations to some black-box models. Their future work will be focused on what [67] has proposed to explore and develop models that are transparent with respect to their decisions. Their work argues about 'right to ask for an explanation from ML-based model' and discuss the future of using intrinsically interpretable models to open black-box detectors.

### 3.4.8 LIME and SHAP-based

The LIME and SHAP-based interpretation use model-agnostic methods to interpret the models. LIME-based methods use linear interpretable surrogates to fit the nearest samples beside a specific data point. Among the process, researchers give noise and disturbance to this data point in order to generate more similar samples. At last, this method can give a local interpretation. In addition, unlike the LIME surrogate, SHAP methods use feature contribution scores to quantify the interpretation outputs and these contribution scores are all signed. For instance, a positive value represents a strengthening effect on the prediction result, whereas a negative value represents a weakening effect on the prediction result.

In [16], the authors addressed that permissions are key factors in Android to protect users' privacy and they have introduced a DL-based detector. Their work used the recent model explanation algorithms such as LIME and SHAP. Furthermore, they mentioned that the model interpretability has only been a loosely defined concept. However, the reason why linear models have more interpretability than CNN or any other DL algorithm has not been proven mathematically. In [48], the authors tried to use LIME to interpret the prediction of malware variants in wild. In general, there are still a few works about using LIME and SHAP-based interpretability towards malware detectors.

### 3.4.9 Reformulation-based

Like gradient-based methods, reformulation-based interpretation methods can also use a gradient, but they will reformulate the whole model by novel approaches such as rewriting the SVM models by pooling neuro networks and maximize the activation layers to print the middle results in DL models. The interpretation results are global due to the complete inspection into models.

In [68], the authors proposed an approach of reformulation one-class SVM as a neural network, which could better be obtained via a deep Taylor decomposition (DTD) to propagate the prediction backward in the neural network towards input features and be suitable for explanation. This investigation has highlighted the tackling with pixel-level anomalies which can be missed by a simple visual inspection. Their work not only focused on images and normal samples, but also focused on interpreting anomalies and context awareness without involving experts' manual analysis in the signature-based malware detection. Researchers in this work took up the aforementioned idea in a simpler setting where no field knowledge is necessary and where it is possible to detect the outcomes on a symbolic level related to an attack from malicious behaviors.

During the whole formulation process, the data point is given as input to the neutralized one-class SVM, and outlier scores are then redistributed using DTD, first on the support vectors, and then propagated further on the input dimensions such as pixels. Utilizing this architecture to highlight the relevant words which are identified as malicious attacks will help to interpret the malware detection.

In addition, the process that converts structures to neuro network can turn some nonlinear regression models into deep neuro networks, which is beneficial for us to use backward explanation and saliency mask to interpret the input features. Undoubtedly, reformulation-based methods outperform the gradient-based methods by a significant margin, and thus the applicable scenarios can be more.

### 3.4.10 Adversarial example-based

Using adversarial examples to interpret detectors is sometimes effective but only gets local interpretation. However, it can be used very often because the interpretation method is lightweight and portable.

In [32], the authors proposed that DNNs are susceptible to tiny crafted adversarial perturbations which are added to all the pixels of the image to craft an adversarial example. Hence, they investigated to generate a new adversarial example attack by flipping pixels of the target image then used the evolutionary method and the bisection method to ignore any model's structure, internal details and training dataset. Their experiment results showed that they could effectively attack MNIST classifier and the ImageNet classifier. Further work will be minimizing perturbation towards features.

In [71], the authors proposed a work for automatic generation of adversarial examples for interpreting malware classifiers, and explored real-world adversarial attacks workflow. Their work proposed new adversarial attacks against the systems in the real world based on randomization and binary manipulation, and it found out that static evasions could also evade the offline dynamic detectors. In addition, their work contributed to explain which features are responsible for evasion.

### 3.4.11 XNN

To use the high performance of NN models with ideal interpretability is a new trend in model interpretability field. The unsolved three issues are: the optimization of resources usage and time consuming, the specific structure waiting to be developed, and the validation of the interpretable structures. Next, we will give a potential architecture for possible use in malware detection interpretability, which is called XNN (explainable neuro network).

XNN [79] is a structured neural network designed to learn interpretable features. Unlike fully connected neural networks, this structure can extract features in a relatively straightforward manner and the results can be displayed, thus to provide a parsimonious explanation of the relationship between features generated from input samples and output results. The structure of XNN provides an alternative formulation of the additive index model as a structured neural network. It also provides a direct approach to fit the model via gradient-based training methods.

The major structural components are: 1) The projection layer, which uses linear activation function. 2) Subnetwork, which learns a potential nonlinear transformation of the input. 3) Combination layer, which calculates a weighted sum the output of the ridge functions.

The input layer is fully connected to the first hidden layer which is called the projection layer. Next, the middle component is the subnetworks, which are used to learn the ridge functions. The combination layer is the final hidden layer of the XNN, which consists of one single node. XNNs use this structure to visualize the univariate, nonlinear transformations learned in training and the projection coefficients. The networks can combine input features used to each ridge function in order to display the most relevant features of the network: the scaled ridge functions and the projection coefficients.

The interpretability of XNN is based on visualizing univariate effects and model recoverability. In practice, XNN can still have a good interpretability on a model without good model recoverability. Besides, XNN can be used as a primary model or a surrogate for more complex models to get high interpretation. However, XNN is a new architecture for model interpretability exploration and it may require more malware detection fields knowledge to acquire a more comprehensive evaluation towards malware detector interpretability.

Nonetheless, the reason we put XNN here is that this model can get high interpretability and high prediction accuracy at the same time due to the nature of neuro network. Instead, traditional ML models like DT, DR and linear regressions can only get high interpretability but low prediction accuracy.

### 3.4.12 The discussions on several newest methods and a critical survey

Although most of the work can be classified into the aforementioned method types, several specific work with ideal performance and interpretability cannot be classified into any of them. For instance, the methodology of XMAL cannot be categorized into any type of the above sections due to that XMAL has used both rule-based and gradient-based techniques. The methodology of LEMNA cannot be categorized also because the related authors used both gradient-based and informative feature-based methods and both of these methods can decide the interpretation results in their work.

In this section, we summarized and discussed the work of LEMNA and XMAL. Next, we discussed a recent critical survey towards interpretability to further explore the future of this field.

In [20], the authors introduce LEMNA, a new method to derive high-fidelity explanations for binary-reverse engineering and malware detection. This framework can approximate the decision boundary through a mixture regression model enhanced by fused lasso in black-box ML-based models. In addition, they point out that RNN and multi-layer perception (MLP) can better predict the sequences from malware features rather than grayscale image processing. They evaluate the interpretation method by first developing the DL-based models to get high prediction accuracy and next by using PCR (positive classification rate) to show the high explanation fidelity that LEMNA can have. Then, their work gives a detailed evaluation experiment by defining feature deduction test, feature augmentation test, and synthetic test. Lastly, they demonstrated how to explore the potential heuristic and point out that LIME and other newest interpretation methods can perform as awful as the feature selection methods due to the sparse input feature vectors that affect in the local decision boundaries.

In [85], the authors presented a novel ML-based approach named XMAL to classify Android malware with high accuracy. With regards to the bias generated from not considering the relevance between each input feature via using gradient-based or reformulation-based methods such as DTD in section 3.4.9, they provided a customized attention mechanism with the MLP model, which consists of attention layer and multilayer perceptron. Next, after choosing the most important feature generated from this framework, they used rule-based filters to produce the natural language descriptions and thus compared them to those coming from the expert analysis. The experiment showed that the prediction is accurate enough to give an interpretation. Lastly, compared with LIME and Drebin, XMAL can obtain better interpretability than the other two methods.

The strengths of this framework are: 1) The evaluation is based on a large number of malwares and benign apps. 2) Using natural language description for the real-time reports from apps and malwares which are easy for us to understand and interpret the prediction outcomes. 3) Using MLP and attention mechanism in order to consider the relevance between dependent input features, which is better than using saliency mask. 4) Producing a customized attention mechanism through a fully connected network to learn the correlation between scalar value features, so as to get rid the necessity of using vector forms to express the elements and targets. The limitations of this framework are that they did not test the effect of using multi-layer attention mechanism and give a comparison and the interpretation features are not comprehensive. Although they can generate the specific features which are important in the prediction compared to the manual analysis, the features are not comprehensive enough for a more complicated malware sample. Hence, the framework requires further exploration.

In [82], the authors gave a survey for explaining black-box models. First, they recapitulated the necessity of interpretability towards valuable data in the system and service. Next, they addressed that interpretability work towards models sometimes can give insight to the input data contributing to the final decisions. They focused on some specific definitions towards interpretability issues, and addressed that high dimensions, the depths of decision trees, and the scales of rule sets can affect the interpretability even though these methods are basically interpretable. In their survey, they summarized the black-box models which have already been interpreted by some methods such as neuro network, tree ensembles, support vector machine, deep neuro network, and nonlinear models. After listing the work, they addressed the importance of model-agnostic methods consisting of the interpretable prediction factor called AGN. Moreover, they gave a list of global interpretation methods including Decision Tree (DT), Decision Rules (DR), Feature Importance (FI), Saliency Mask (SM), Sensitivity Analysis (SA), Partial Dependence Plot (PDP), Prototype Selection (PS), and Activation Maximization (AM). The core contribution of their work is that they summarized the related work towards solving the model explanation problem, outcome explanation problem, model inspection problem and the transparent box design problem via the evaluation indicators like data types, explanators covering the above 8 types of methods and the basic information related to the experiments of these work like examples, code, and datasets. Lastly, the key discussions existing in their survey include two aspects: one is that whether we have an exact definition of interpretability, whereas the other is that whether we can use interpretation methods to discover the hidden features existing in the models. As long as we can discover the hidden features, we can accomplish more complicated goals like splitting the malware code automatically by ML-based models, recognizing the edges of malware functions, and predicting the malware variants with comprehensive interpretation.

In general, although malware detection can get interpretability to some degree via behaviors description such as the work in [85], using the state-of-the-art techniques such as attention mechanism and common interpretation methods can still have potential for researchers to further explore their effectiveness. Moreover, model and features visualization can also be as crucial as using the natural language to describe the malware behaviors. We note that these image-based methods still have a long way to go because the saliency mask processing the image-based features always meet with barriers like vanishing gradient and not considering the relevance among input features. In [82], the authors provided four key problems related to opening the black-box models, but the current researches have not yet been concerned a lot about these four issues when interpreting the models. Besides, the accuracy and interpretability, the number of the features and interpretability, and the nature of features and interpretability, all these three pairs of relationships need further discussion and research in the future.

### 3.4.13 Summarizing and comparison in details

As mentioned above from Section 3.4.1 to Section 3.4.12, all of the work can have strengths and limitations, and some of the work are not fully concerned about detection interpretability. In fact, the present malware detection models based on ML focus more on model performance, and there are a small number of works completely focused on the model interpretation.

**Table 6** summaries the experiment performance and interpretability results of these work. **Table 7** describes the strengths and limitations of these work.

We note that: DNN: DL neuro network, CNN: convolutional neuro network, CME: covariance multiplication estimator, FDA: fisher discriminant analysis, SVM: support vector machine, FNN: first near neighbors, ANN: all nearest neighbors, WANN: weighted all nearest neighbors, KMNN: k-medoid based nearest neighbors, DT: decision tree, MLP: multilayer perceptron, RF: random forest, NB: Naïve Bayes, J48: a kind of DT algorithm for use, EMBER: an open-source ML-based classifier using a tree-based classifier model, namely LightGBM, to detect malware, ClamAV: an open-source signature-based antivirus engine utilizing hash-based and rule-based signatures to detect malware.

As is illustrated in **Table 6**, from the Reference [82], we can know that the interpretability work must be set based on the high prediction accuracy, and the data types for interpretation are classified into IMG, TAB and TXT. Among them, TAB can represent a classical dataset and share the common features, whereas IMG and TXT can be used for black-box images input and model building. Besides, we show the specific data sources and numbers so as to give the related situation performance. Next, we give the exact method type to each work by the taxonomy in Section 2.4 in order to validate the correctness of this taxonomy. Because if we can classify any work of all into two or more types at the same time, then we should doubt the correctness of taxonomy in Section 2.4. However, we can find that each work can be categorized into a single type. Next, we use black-box targets, explanators and interpretation fidelity (note that fidelity comes from Section 2.3) to make the evaluation complete. **Table 7** summarizes the advantages and limitations existing in the work.

### 3.5 Evaluation system using comprehensive interpretability score

In [82], the authors note that currently there is no valid methodology to quantify the levels of interpretability towards models. In our survey, we first try to give different weight as interpretability score for malware detectors as our evaluation system. Our aim is to find the most potentially interpretable methods and give insight into quantifying the evaluation in the future work.

#### 3.5.1 Evaluation methodology

We use the attributes described in Section 2.2 to be our indicators for the following evaluation. Four attributes together have 11 aspects used as factors in our evaluation. We add 2 to feature summary statistic, 2 to feature summary visualization, 2 to model internals, 1 to data point, 3 to intrinsically interpretable model, 2 to model-specific, 2 to model-agnostic, 1 to local, 2 to global, 2 to intrinsic, 2 to post hoc.

The reasons why we give these weights **as follows**. Note that the weight value is set by our own experience.

First, intrinsically interpretable models can be interpreted due to their interpretable structures such as DT, linear-regression models and DR etc. In fact, these models do not need interpretation methods unless their depth or scales become complicated enough [82]. In most cases, these interpretable models can give global and steady interpretation results without any special interpretation methods. Hence, we give the intrinsically interpretable models methods the weight of 3 as the highest score.

Next, feature summary statistic and feature summary visualization can have outcome interpretation. For instance, References like [9][12][85] use malware grayscale interpretation, rule-based methods and natural language description via the rule-based filters relatively. The interpretability of feature summary cannot be as high as the intrinsic model interpretability because their outcome explanation can sometimes only lead to local interpretation, which means the interpretation results can only suit the specific situation and might fail in other cases. Hence, we give the weight of 2 to these two factors.

- **Table 6**. Summary of methods for interpreting malware detection problems: results performance (prediction, attacking), interpretability (explanator, fidelity, data type, method type)

| Basic information | | | Experiment performance | | | Interpretability | | | |
|---|---|---|---|---|---|---|---|---|---|
| Ref. | Input | Year | Black Box as targets | Data sources/No. of samples | Results for classification | Explanator | Fidelity | Data Type | Common type |
| [9] | Grayscale images | 2018 | DNN (pre-trained on natural images and retrained) | 9458-malware samples: 25 families | Accuracy-99.25%,98.3%,99.67% | Super-pixel representation | × | IMG | Neuro network interpretation |
| [10] | Markov image | 2020 | CNN | 10,868 labeled samples from the Microsoft:9 families; 4020 android malware samples: top 10 families | Accuracy-99.264%,97.364% | n/a | × | IMG | Neuro network interpretation |
| [11] | Informative variables | 2020 | CME, FDA, SVM | Linear data with a set of 485 input features; non-linear data; MNIST | n/a (not used for classification) | CME; visualization | √ | TAB/IMG/TXT | Model-specific |
| [41] | Obfuscators and detectors | 2020 | Any black-box models | 1260 malware samples; 1260 benign samples | n/a (not used for classification) | Feature obfuscation (data points) | √ | TAB/TXT | Example-based |
| [73] | APIs, permissions, intents | 2020 | FNN, ANN, WANN, KMNN | 118,505 Drebin samples; 1200 Genome samples; 11960 malware samples and 16800 benign samples from Contagio | Accuracy-99.31%(ANN), 99.33%(FNN), 99.31%(WANN), 99.28%(KMNN) | n/a | × | TAB | Example-based |
| [76] | Opcode | 2019 | KNN, SVM, MLP, Ada-Boost, DT, RF | Virus Total 2017 and number is unknown | Accuracy-99% | n/a | × | TAB | Example-based |
| [12] | Images | 2020 | CNN, RBF-NF | MNIST:50000 training samples and 10000 testing samples | Accuracy-99.16% | RBF-NF | √ | IMG | Neuro network interpretation |
| [16] | Permissions | 2020 | CNN for NLP | 77000 descriptions from Google Play | Accuracy-71% to 93% for recognizing the usage of dangerous permissions | LIME | √ | TAB/TXT | Model-agnostic |
| [48] | Free-form text and words | 2016 | NLP models including NNs, SVM | n/a | Accuracy-94% | LIME/SP-LIME | × | TAB | Model-agnostic |
| [25] | IF-THEN rules | 2019 | n/a | n/a | n/a | Rules and feature description | × | TAB/TXT | Model-specific |
| [18] | Import feature, strings, PE header opcode | 2019 | Interpretable feed-forward neural network | 4361 benign samples and 6096 malware EXE/DLL samples from MalShare and Virus Total (2018) | Accuracy-98.5% | IFFNN | × | TAB | Neuro network interpretation |
| [72] | Packers attributes | 2020 | Packers classifiers | 3956 unpacked benign samples and 3956 packed malware samples; 107471 samples for training and 46059 samples for evaluation | Accuracy-91.14% | Feature selection | × | TAB | Model-agnostic |
| [28] | CSV | 2020 | RF | Windows 7: 95191 malware samples, 953384 benign samples; 146 samples for NODES | Accuracy-100% (Ransomware, Spyware, Backdoor, Bit Coin Miner, Process Injector and virus), 93%(Trojan) | NODES | √ | TAB/TXT/IMG | Model-specific |
| [69] | API calls from dynamic analysis | 2020 | NB, KNN, SVM, DT, RF, Boosting | 6434 benign samples and 8634 malwares from windows 7, XP-SP3, and 10. | Accuracy-99.1% | Model internals pruning | × | TAB | Model-specific |
| [70] | Permission requests and API calls | 2020 | RF, J48, RT, KNN, NB | 27891 Android apps for malware datasets | F1-score-94.3% | Feature description | × | TAB | Model-specific |
| [74] | API calls, permissions | 2020 | Semi-supervised learning models | n/a | n/a | n/a | × | TAB | Example-based |

| | | | | | | | | |
|---|---|---|---|---|---|---|---|---|
| [75] | Sensitive permission requests and app textual description | 2020 | NLP models | MAL_AMD:24553 samples, AndroZoo, Contagio, Other sources: DREBIN, VirusShare, AndroidPRAGuadDataset, MamaDroid and DroidSpan dataset, etc. | Accuracy-98.72%, 96% | Textual description | × | TAB | Neuro network interpretation |
| [64] | Strings | 2018 | SVM with RBF kernel; RF for testing whereas any black-box models for potential interpreting | Drebin:121329 benignware samples, 5615 malware samples | Accuracy-99% or more | Gradient-based method | √ | TAB | Neuro network interpretation |
| [68] | Images and input variables | 2020 | SVM | MNIST, CIFAR-10, Detox dataset | n/a | Reformulation and gradient-based method called DTD | × | TAB/TXT/IMG | Neuro network interpretation |
| [71] | Manipulation actions sequence from adversarial examples | 2020 | EMBER, ClamAV | 1000 malware samples for attacked | 24.3%-41.9% for evasion by changing one byte | In-place randomization and code-agnostic manipulations | × | TAB | Model-agnostic |
| [32] | Images | 2020 | Black-box models | MNIST, ImageNet dataset, CIFAR-10 | Excellent attack generation results | Generation of attacks to gain insight from interpretation | × | IMG | Example-based |
| [79] | n/a | 2018 | XNN | n/a | Explainable neuro network | Appropriate regularization | √ | TAB/TXT/IMG | Model-specific |
| [85] | n/a | 2020 | XMAL | 5560 malware samples by Drebin: top 16 families | Accuracy-97,04% | Attention mechanism into MLP | √ | TAB | Neuro network interpretation |
| [20] | Binary sequences, opcode, feature sequences from the PDF malware and reverse engineering | 2018 | MLP-based models, and binary reverse-engineering | 4999 malware samples from PDF files, and 5000 benign files | Accuracy-98.64% | LEMNA | √ | TAB/TXT | Model-agnostic |

- **Table 7**. The discussion of methods from Reference (advantages and limitations)

| Ref. | Advantages | Limitation |
|---|---|---|
| [9] | It can interpret the grayscale images to some degree, and show the most related features contributed to the trained model. | There is no fidelity for interpretation when features come more and no trust score to quantify the level of interpretability. In addition, it is not proved that it can interpret the adversarial and obfuscated samples. |
| [10] | It can convert the binary files into **Markov** images which is better than grayscale images for not losing some bytes and better for interpretation. | There is no specific interpretation on Markov images and more detailed exploration. Besides, the interpretability work on Markov images requires more. |
| [11] | It can use a FS method and estimate the interdependence of the features. Besides, they use visualization to convert features relevance into graphs and finally rank the features. Lastly, the interpretation work is comprehensive. | There is some fidelity but lack of detailed verification. Besides, whether this interpretability framework can be applied on huge malware variants requires further exploration. |
| [41] | They obfuscate various Android malware samples to affect the detectors and determine the related features. Besides, they show that there are various features contributing to a number of detectors. | They need further explore the robustness of HMM-based, SVM-based, and DL-based malware detectors and interpret them by obfuscating the features. |
| [73] | They use Hamming distance to detect malware and compare several detectors. They present a more accurate experiment and the potential for interpretation based on APIs, permissions, and intents. | There is no specific estimation for interpretability even though the models and detectors can have potential for understanding. |
| [76] | They use sequential pattern mining technique to detect MFP from opcodes, and give insight and potential for interpreting. | There is no specific estimation for interpretability even though the models and detectors can have potential for understanding. |
| [12] | They use Fuzzy logic and interpret the structure's rule base with linguistic capability. | There is no description for feature maps generated from RBF-NF for malware detection even though it has strong potential for image-based malware samples. |
| [16] | They use CNN to process the NLP problem in malware detection and interpret the dangerous permission and usages related to words. | They need a further exploration on fidelity and feature selection towards interpretation. Besides, a more comprehensive interpretation description and compared experiment need to be set up. |
| [48] | They use LIME and SP-LIME to interpret the NLP problems. | They need more exploration on malware detection interpretability and the effectiveness on LIME for interpreting due to that there is already a better choice like LEMNA. |
| [25] | They use IF-THEN rules to generate the understandable description for malware analysis rather than the raw ML-based features and further explore the emerging malware threats. | They need to further explore the fidelity of model interpretability. Besides, the anti-adversarial malware can give insight into interpretation rather than describing the correct decision. With regard to this, they point out many discussions but lack practice. |
| [18] | They use different feature attributes to quantify the detection interpretability like: the impact score of assembly code, number of PE imports, major operating system versions, frequency of the string 'Sleep', and the frequency of the string '.data'. Next, the I-MAD can give both high prediction accuracy and understandable interpretability. | The fidelity of the interpretation requires further verification and the interpretation factors are not defined systematically. |
| [72] | They use tremendous number of static features from samples and explore the effect of the packers. The interpretation and comparison for packers can give insight to the wild malware variants. | There is no specific interpretation evaluation towards malware detection systematically and only the data points like packers and obfuscation can give feature contribution, which lacks the exploration towards model interpretability. |
| [28] | The NODES is light, accurate and interpretable. Besides, the explanation can be understood easily by analysts which can give insight for further detection. | The datasets for NODES are only 146 malwares and malware samples run in the virtual machine which cannot avoid the problem of 'VMaware' malware. In general, a more comprehensive detection towards more complicated malware variants requires exploration. |
| [69] | They use dynamic analysis to detect the malware by APIs and prune the model internals to obtain the best performance. | They lack systematical analysis towards interpretability through the process of model pruning. |
| [70] | They use API calls and permissions in Android apps for malware detection and give insight into the transformation for these features in order to get interpretability to some degree. | They lack systematical interpretability evaluation and mainly focus on the process of detection. |

| [74] | They use semi-supervised learning models to get accurate prediction and give insights into malware variants in the wild. | They lack interpretability evaluation in details. |
|---|---|---|
| [75] | They use textual description to give insight into malware detection by permissions and graphical user interface inputs. | They lack detailed interpretability evaluation systematically and thus cannot ensure the fidelity of interpretation. Besides, the textual descriptions can be understandable, but only give the local interpretation. |
| [64] | They give systematical exploration towards malware detection interpretability in local and global aspects. Furthermore, they use gradient-based methods to give mean relevance scores for malware samples in different models: SVM, SVM-RBF, and RF. Besides, they visualize the relevance features. Their explainable approach can help to understand possible vulnerabilities of learning algorithms to well-crafted evasion attacks along with their transferability properties. | They need to further explore the usage of different surrogate models to affect their interpretation outcomes. |
| [68] | They present a novel approach using DTD for saliency masking and convert the SVM into neuro network by a process that converts structures to neuro networks. Hence, the method can either interpret the images or the datasets from language texts, which outperforms the traditional DTD processing towards only the images. | They lack the discussion on discovering the anomalies from malware textual description in order to explore the inner nature of adversarial malware samples which is better for interpretation. Besides, they haven't considered the relevance between the feature inputs, which cannot be proved as effective as attention mechanism. Lastly, the DTD and the process that converts structures to neuro networks turn interesting, but they lack interpretability evaluation in details for malware samples and haven't focused on the drawbacks. |
| [71] | They first prove the drawbacks hidden in the commercial anti-virus software with dynamic and static engines. Besides, they generate the adversarial malware samples to boost the interpretability in the malware detectors. In general, they design a generalized, systematic, automatic framework to perform realistic adversarial attacks on real-world malware detectors. | They lack the systematical evaluation towards interpretability. Furthermore, they lack the relevance between the generalization process and the detection interpretability which can be further discussed in the future. |
| [32] | They use a newly developed black-box attack method based on the evolutionary and the bisection method. Besides, the attacking model does not require confidence for prediction but only the final class labels. | They merely generate the attacking framework but lack the interpretability evaluation. Besides, using adversarial data points can lead to local interpretation which need further discussion. |
| [79] | They design a new explainable neuro network different from the fully-connected network, and use regularization to interpret the relationships between features and outcomes. | They haven't applied this model to the malware detection, and the actual interpretability evaluation score requires further discussion and testing. |
| [85] | They design a framework consisting of attention mechanism and MLP, and generate the malware feature description for understandable interpretation with high fidelity by comparing the descriptions with those generated from experts. | The XMAL lacks the multi-layer attention mechanism embedding, and the features for interpreting models are not comprehensive enough. |
| [20] | They design a new interpretation method called LEMNA which is enhanced by fused lasso and mixture regression models. LEMNA can give the interpretation involving the relevance between input features and dependencies. Besides, LEMNA can perform better than LIME according to the metrics and experiments by PCR factors and pruning the hyperparameter. | They lack the fidelity of interpretation and give a little attention on a comprehensive interpretability evaluation. Besides, it is better to describe the interpretability generated from features. |

Moreover, model internal, model-specific, global and intrinsic, these four factors can only refer to the specific models. They cannot have the model-agnostic nature like LIME, but the interpretability towards one single model can be ideal. Besides, these factors can give insight into data features for further exploration due to that they are more focused on the inner structures of models. Hence, we also give the weight of 2 to these factors.

In addition, model-agnostic and post-hoc attributes are two factors that get rid of the specific models because both of them are only concerned about the relations between input features and outcomes. For instance, model-agnostic methods like LIME can use some interpretable surrogate models to understand the black-box prediction decisions, which obviously skip the original black-box inner structure. Since these post hoc methods can sometimes perform as well as model-specific methods in the experiments, we also give the weight of 2 to these factors as the same with the above four factors.

Lastly, some data points like counterfactual samples affecting the trained models and using different data points to interpret the prediction often lead to local interpretation. It is hard for this method to obtain the comprehensive and global malware prediction interpretability. Therefore, we give the weight of 1 to the local and data points.

In general, although we cannot specify the exact scores through the current experiments due to the lack of experimental validation, we can still set these weight values. Because the intrinsic model can surely get the best interpretability, and except for the data points and local these two factors, all the other factors turn to the same weight of 2. It can be a good choice because all factors cannot be measured in common unless we go deep into the specific experiments. Hence, we give weight to approximately quantify the interpretability score of these work from Reference.

Now we will show our evaluation method. Let the $k^{\text{th}}$ type of work model interpretability score be denoted as $I(k)$, i.e. $w(i)$ is the $i^{\text{th}}$ interpretability indicator weight, then we have the cumulative score of all the indicators existing in the $j$th work which is the comprehensive score $W(j)$.

$$W(j) = \sum_{i=1}^{n} w(i) \tag{1}$$

Next, we will get the interpretability score of the specific type of work, given the fact that number of work in this type is $m$.

$$I(k) = \frac{1}{m} \sum_{j=1}^{m} W(j) \tag{2}$$

### 3.5.2 Evaluation results

We give the comprehensive score for each research as shown in **Fig.6**. In general, we can see from **Fig.6** that the work of highest score is [28], whereas the lowest score work is [10]. From the literature, we can know that the highest score work involves with RF interpretable model, and it can get the global interpretation towards malware detection. However, the lowest score work mainly focuses on malware classification with interpretable features, which merely lead to local and model-agnostic interpretation. The histogram indicates that we should use interpretable models to do the accurate prediction and focus more on intrinsically model interpretation rather than finding correlate in the feature space.

In addition, according to the taxonomy we provide in Section 3.4, we then calculate the average score of all work in each type as the interpretability score of different detector interpretation work as illustrated in **Fig. 7**. We observe that there comes with the highest score of 13 towards reformulation-based methods. Furthermore, reformulation-based methods give insight into black-box model so as to get rid of the local interpretation on the outcomes.

From the aspect of feature extraction, we can see that dynamic-based methods can be interpreted better than the static-based methods due to that the dynamic features always refer to real-time behaviors, whereas the static features are always redundant, noisy and obfuscated. From the feature processing aspect, we can see that image-based methods cannot get high interpretability even though they are intuitive for researchers to observe, because we cannot truly understand the meaning of the grayscale images. In addition, the rule-based methods can perform well but not better than gradient-based, dynamic-based and reformulation-based methods. Because rules can help understand the results but lack the dependability compared to using dynamic analysis and using some novel approaches which have been proved mathematically to open the inner structure of black-box models. Lastly, the formative-feature-based methods can get high interpretability score because they use multimodal features to make interpretation results complete.

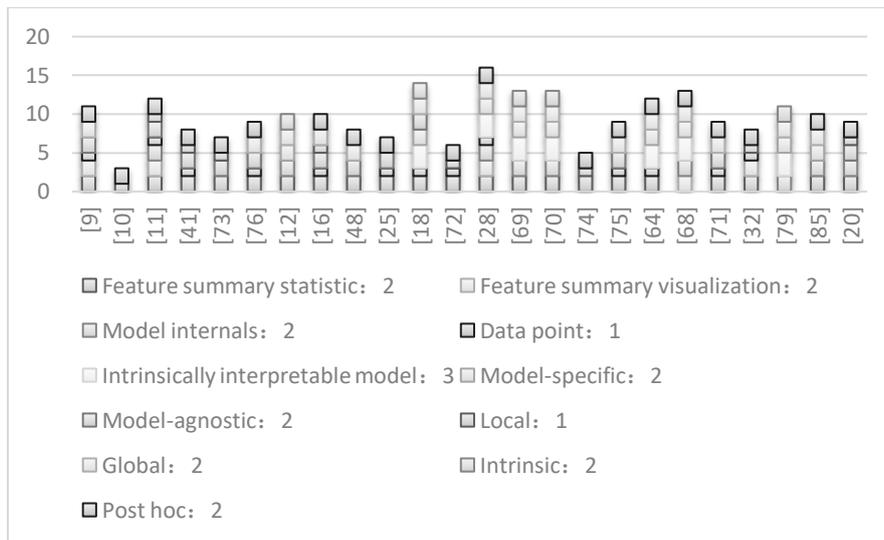

**Fig. 6 Comprehensive score of the main state-of-the-art malware detection researches and the composition of the score**

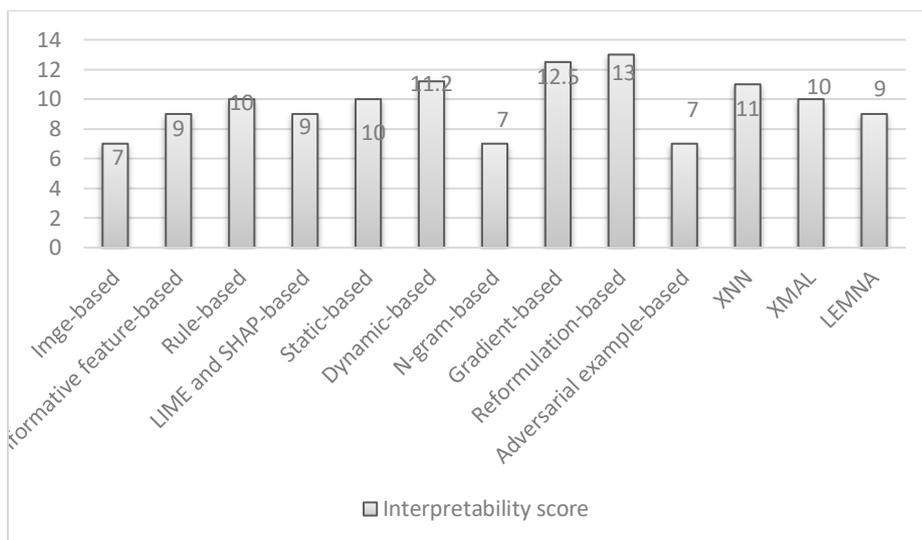

**Fig. 7 The state-of-the-art malware detector interpretation methods evaluation and comparison: interpretability score after averaging process**

From the aspect of model internals, we can see that LIME, SHAP, gradient-based, reformulation-based, and LEMNA methods can all have better interpretation results than on the feature extraction aspect. Next, the intrinsic model XNN and the comprehensive method called XMAL can interpret the prediction results well because the inner structures of them are interpretable to some degree (note that they cannot be as interpretable as the DT, DR and so on).

### 3.5.3 Open research issues towards the evaluation in this survey and the direction of effort towards future interpretation architecture

Although we want to use model interpretability indicators described in Section 2.3 to give a more comprehensive evaluation and comparison towards all these works, it is difficult to evaluate them in such multiple dimensions because these work lack the specific indicators **more or less** via experiments and thus we cannot just measure them by giving different values to these indicators. In fact, we only consider the fidelity in Section 3.4.13. We hope that the future interpretability work can be evaluated by all indicators described in Section 2.3, and thus researchers can get a more comprehensive summary.

In addition, the tradeoff among prediction accuracy, interpretability, as well as the nature and number of input features also matter very much in this field. Hence, we provide an open research issue of whether we can have a more comprehensive survey of model interpretability both with model accuracy, so as to give an ideal developed version of models in the future. Besides, all these works also needs to be considered with time and resources consumptions. Lastly, a reasonable quantitative level measurement system requires exploration to evaluate the indicators in Section 2.3 and the completeness of these indicators needs to be proven in the future. Note that in Section 3.4.13 we only quantify the attributes of interpretation methods by own experience and a more comprehensive evaluation involving mathematical proven process needs explored in the future work.

### 3.6 Conclusion

By the detailed depiction of recent work on ML-based analysis models and their interpretability, we can know that feature selection and correlated regulation between input features and outcomes are categorized into many parts. Although there are a number of approaches to resolve ML model interpretability, malware detector interpretability should be different due to the samples which are more complicated than any other field dataset. Expert knowledge can be applied to boost model processing interpretability and reduce resource consumptions. However, this field lacks fully automated models based on interpretable workflow derived from raw datasets and small-scaled samples. Besides, a quantitative level evaluation towards interpretability experiments need to be explored. Lastly, an automated system for interpreting DL work on malware detection has not been proposed, instead there are a number of manually interpreting work on malware detection which is always after detection results. A balance among accuracy, efficiency and interpretability still has a long way to go for optimization.

While keeping the interpretability of these multimodal approaches, we must accelerate the development of the classifiers for automated detection and apply these technologies for more polymorphic sample datasets. **Fig. 8** shows the possible future work which may require further exploration as well as the current motivation on improving ML model interpretability.

As is illustrated by **Fig. 8**, model interpretability also requires automated system to take place of manual work, and multimodal approaches can be improved and found by the insight given by these interpretation results. Besides, optimizing computation resources usage and accelerating the running speed should always be considered in every stage so as to improve the portability and alleviate the algorithm complexity.

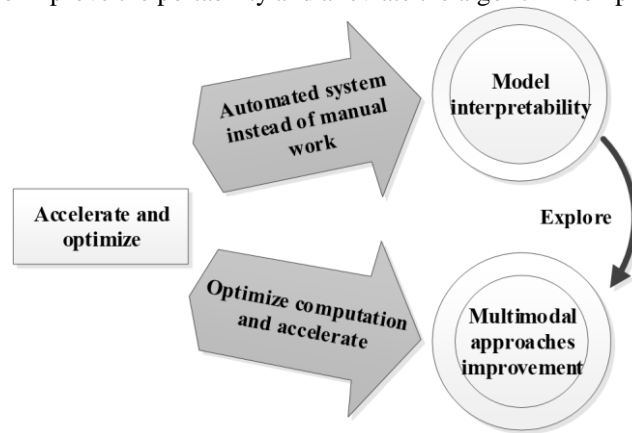

**Fig. 8 Research prospects in model interpretability**

# 4 Summary and prospect

To our best knowledge, we are the first to provide the main taxonomy of malware detector model interpretability. Most of the ML-based models that embedded in the malware detectors are the black-box models requires to be interpreted for its prediction results and entire workflow. Our survey includes two parts. The first part introduces the common interpretation methods with regards to their metrics, attributes, evaluation indicators, and an existing taxonomy from Reference. Besides, we compare these methods in groups via several indicators. The second part reviews the malware detection interpretability solution logic and describes the state-of-the-art malware detection models interpretability researches via dividing malware detection interpretation methods into several categories. Besides, we discuss several novel interpretable architectures like XNN, XMAL, and LEMNA. Then we discuss one current important survey to give more descriptions towards black-box interpretation. Next, we compare the work based on their experiment performance and interpretability results. Lastly, we evaluate the interpretability score of these work by giving different weight to the common interpretation methods attributes described in Section 2.2.

Although work on ML model interpretability has got a great progress, the interpretation work towards malware detectors based on black-box models is unable to be evaluated and compared in a systematic way. Given the above issues, we firstly try to address the solution logic of malware detection interpretability and provide a new taxonomy to classify the state-of-the-art interpretation methods in the malware detection field. Moreover, the survey discusses several critical works of exploring interpretability and provide a novel approach to alleviate the issue of being unable to quantify the interpretability evaluation indicators.

From the detailed overview and evaluation through these works, we can know that the black-box based malware detector models needs to be given more investigation on computational cost and their interpretation performance under the circumstances of more complicated environment based on polymorphic malware variants. In addition, a rigorous mathematically prove quantitative system for model interpretability needs to be studied and provided.

As a future work on this field, in addition to more detailed manual expert work after automated ML-based detectors operate, we will take more automated interpretability models into consideration to strengthen the trust on a more complex black-box malware detector. In addition, researchers should focus on protecting malware detection system as well as the entire workflow between prediction and interpretation phases.

## Availability of data and material

Not applicable.

## Funding

Not applicable.

## Acknowledgements

Not applicable.